\documentclass[twoside]{article}
\usepackage{times}
\newcommand{\be}{\begin{equation}}
\newcommand{\ee}{\end{equation}}
\newcommand{\bea}{\begin{eqnarray}}
\newcommand{\eea}{\end{eqnarray}}
\begin{document}

%%%%%%%%%%%%%%%%%%%%%%%%%%%%%%%%%

\vspace*{-2cm}

\begin{center}
{\footnotesize 
Proceedings of the First International Workshop on {\em Symmetries in
Quantum Mechanics and Quantum Optics}, A. Ballesteros, {\it et al} (Eds.) 
Servicio de Publicaciones de la Universidad de Burgos (Spain), p.~285-299. 
Burgos, Spain (1999).}
\end{center}
\vskip0.5cm 
%%%%%%%%%%%%%%%%%%%%%%%%%%%%%%%%%%

\begin{center}

{\LARGE{\bf{On the factorization method in}}}

{\LARGE{\bf{quantum mechanics}}}

\vskip1cm

{\sc J. Oscar Rosas-Ortiz}
\vskip0.5cm
{\it
Departamento de F\'{\i}sica Te\'orica,
Universidad de Valladolid\\
E-47011 Valladolid, Spain\\ [0.1cm]   
and\\ %[0.1cm]
Departamento de F\'{\i}sica,
CINVESTAV-IPN, A.P. 14-740\\
07000 M\'exico D.F., Mexico.
}
\end{center}
\bigskip
\vskip1cm

\begin{abstract}
\noindent
New exactly solvable problems have already been studied by using a
modification of the factorization method introduced by Mielnik. We re\-view
this method and its connection with the traditional factorization method. 
The survey includes the discussion on a generalization of the
factorization energies used in the traditional Infeld and Hull method.

\end{abstract}

%%%%%%%%%%%%%%%%%%%%%%%%%%%%%%%%%%%%%%%%%%%%%%

\section{Introduction}

Since the begining of the Quantum Mechanics (QM) there exists only a
narrow set of exactly solvable physical problems. This set includes the
well known harmonic oscillator and hydrogen-like potentials, the Morse
potential, the square well potential and a few others. The study of such
problems has been traditionally done by using the method of orthogonal
polynomials and the factorization method. In principle, it is no matter
the method used in the construction of the analytical solutions to the
Schr\"{o}dinger equation, however, the factorization (introduced by
Schr\"{o}dinger \cite{S40} and Dirac \cite{D35}) avoids the use of
cumbersome mathematical tools and it has been succesfully applied as an
elegant approach in solving essentially any problem for wich there exists
exact solution. This is clear from the classification and study of the
factorizable second order differential equations given by Infeld and Hull
\cite{IH51}. It is remarkable that, although the factorization method was
developed to solve the eigenvalue problem related with the
time-independent Schr\"{o}dinger equation, it is also a very power\-ful
tool in the study of the recurrence relations obeyed by special functions
\cite{M68}. 

Until the begining of the 1980's decade, it was a very common opinion that
the method was completely explored. The change of mentality became after
the seminal work of Mielnik \cite{M84}. The fundamental idea introduced by
Mielnik in such paper is to consider not the particular but the general
solution to the Riccati type equation connected with the
Schr\"{o}dinger-Dirac-Infeld-Hull (SDIH) approach. Hence, the Mielnik's
method takes into account an important term of the factorization operators
unnoticed by the traditional SDIH method. The first application of this
modified method was developed in the study of the harmonic oscillator by
Mielnik himself, giving rise to the construction of a {\it one}-parameter
family of new exactly solvable potentials, different than the harmonic
oscillator potential but having the same spectrum as it.  Early on, the
Mielnik method was applied by Fern\'andez in the construction of a {\it
one}-parameter family of new exactly solvable radial potentials,
isospectral to the hydrogen-like radial one \cite{F84}. In the same
year Nieto gave the links between this modified SDIH method and the
supersymmetric (SUSY) approach to QM \cite{N84}. Its connection with
Darboux transformation \cite{D882} has been discussed by Bagrov and
Samsonov \cite{B96} (see also \cite{A85}), and some links with the inverse
scattering method have been given by Zakhariev {\it et al.} \cite{Z96}. 
By now the Mielnik method has been maturing for one and a half decade, and
far from becoming a closed chapter in solving the Schr\"{o}dinger
equation, it is continuously reviewed and applied in diverse prospects
[12-16]. For more historical links and related approaches see the
interesting account given by Rosu in this volume.

In this contribution we shall present a general survey of the
factorization method. Our main interest is to state clearly the
differences between the SDIH and the Mielnik methods and, at the same
time, to give the introductory material allowing their application in the
derivation and study of new exactly solvable problems. In Section 2 we
will review the SDIH method. Section 3 deals with the Mielnik method.
Finally, in Section 4, we shall review the further generalizations of the
Mielnik's approach recently introduced in \cite{F97,R98}. Each section
contains the corresponding harmonic oscillator and hydrogen-like
potentials as particular cases.

%%%%%%%%%%%%%%%%%%%%%%%%%%%%%%%%%%%%%%%%%%%%
\section{Traditional factorization}

The Schr\"{o}dinger equation for a time-independent one-dimensional system
is given by the usual eigenvalue equation
\be
H \psi (x) = \left[ -\left( \frac{\hbar^2}{2m} \right) \frac{d^2}{dx^2} +
v(x) \right] \psi (x) = {\cal E} \psi(x),
\label{1}
\ee
where the potential $v(x)$ can be a singular ({\it e.g.} the hydrogen-like
potential) as well as a nonsingular (like the free particle or the
harmonic oscillator potentials) real function, and the wave function $\psi
(x)$ must be {\it square-integrable}, that is, the integral $\int \vert \psi
(x) \vert^2 dx$ is finite. One often uses wave functions which are
normalized, such that:
\be
\int \vert \psi (x) \vert^2 dx = 1,
\label{2}
\ee
equation (\ref{2}) is usually called the {\it normalization condition}.

As is well known, any physical problem must state not only the
differential equation which is to be solved but also the boundary
conditions which the solution must satisfy. Moreover, it is remarkable to
notice that the determination of solutions satisfying of the boundary
conditions is often as difficult a task as the solving of the differential
equation. At the present state, it is apparent the essential role played
by the normalization condition (\ref{2}) in quantum theory because it
represents the strongest {\it boundary condition} to be satysfied by the
solutions to (\ref{1}).

The spirit of the factorization method, as introduced by Schr\"{o}\-din\-ger
and Dirac, is to write the second order differential operator $H$ in
(\ref{1}) as the product of two first order differential operators $a$ and
$a^{\dagger}$, plus a constant $\epsilon$.  The form of such
operators depends explicitly on the particular potential $v(x)$ we are
dealing with, and implicitly on the {\it factorization energy} $\epsilon$,
as we shall see below. We are going to sketch now a brief survey of the
SDIH method.  In order to eliminate a cumbersome notation let us rewrite
the Schr\"{o}dinger equation (\ref{1}) in the dimensionless form
\be
\left[ - \frac{d^2}{dx^2} + V(x) \right] \psi (x) = E
\psi(x),
\label{3}
\ee
where $V(x) \equiv 2 m v(x)/\hbar^2$ and $E \equiv 2 m{\cal E}/\hbar^2$. 
The factorization method aims to find two operators
\be
a = \frac{d}{dx} + \beta_p(x), \quad \quad
a^{\dagger} = - \frac{d}{dx} + \beta_p(x),
\label{4}
\ee
such that the Hamiltonian
\be
H =  - \frac{d^2}{dx^2} + V(x)
\label{5}
\ee
can be written as
\be
H = a^{\dagger} a + \epsilon.
\label{6}
\ee
It is a matter of substitution to show that $\beta_p(x)$ has to satisfy
the following Riccati equation: 
\be
-\beta'_p(x) + \beta_p^2(x) = V(x) - \epsilon.
\label{7}
\ee

The SDIH method takes into account only the particular solution to the
Riccati equation (\ref{7}), this is the reason why we have labeled the
$\beta$-functions, from the very begining, with the subindex ``$p$''. In
the following table we reproduce the Infeld-Hull results for the harmonic
oscillator as well as for the hydrogen-like potentials. The traditional
representation for the harmonic oscillator is recovered from (\ref{4}) and
(\ref{6}) by take $H \rightarrow H/2$, $a \rightarrow (1/\sqrt{2}) a$, and
$a^{\dagger} \rightarrow (1 /\sqrt{2} )  a^{\dagger}$. 
\begin{center}
\begin{tabular}{||c|c|c||} 
\hline
&&\\ [-1ex] 
Potential & Factorization & Particular solution to\\[1ex]
          & energy        & the Riccati's equation\\[1ex]
\hline 
&&\\ [-1ex]
$V(x) = x^2$  &  $\epsilon = 1$  &  $\beta_p(x, \epsilon) = x$ \\[2ex]
              & ``modified method'' \cite{IH51} &    \\[2ex]
\hline
&&\\ [-1ex]
$V_l(r) -\frac{2}{r} + \frac{l(l+1)}{r^2}$  & $\epsilon = -\frac{1}{l^2}$
&  $\beta_p (r, \epsilon) = \frac{l}{r} - \frac{1}{l}$  \\[2ex]
              &      for a fixed $l$   & for a fixed $l$ \\[2ex]
\hline 
\end{tabular} 
\end{center}
Let us remark that, in general, given any $\epsilon \in {\bf R}$ for which
there exists a solution to (\ref{7}), there is a factorized expression for
$H$ as it is given in (\ref{6}). However, it is worthwhile noticing that,
because the Riccati equation (\ref{7}) is a non-linear differential
equation, it is not, in general, integrable by quadratures. It therefore
defines a family of transcendental functions \cite{I26}, and the search of
an adequate factorization energy permitting the construction of the
$\beta_p$-functions is not an easy task.  In fact, Infeld and Hull have
reported only one factorization energy for each potential $V(x)$, hence,
they have given the first step in the systematic construction of exactly
solvable problems in Quantum Mechanics.

Now, let us take the product (\ref{6}) in the opposite order and, by using
(\ref{4}) and (\ref{7}), we get
\be
a a^{\dagger} + \epsilon = H + 2 \beta'_p (x).
\label{8}
\ee
In a first view, one might be tempted to belive that equation (\ref{8})
brings out a new Hamiltonian $H + 2 \beta_p' (x)$, but it is not always the case,
as we are going to see in the next sections.

%%%%%%%%%%%%%%%%%%%%%%%%%%%%%%%%%%%%%%%%%%%%%%%%%%%
\subsection{The harmonic oscillator revisited}

As is well known, the solutions to (\ref{3}), with $V(x) = x^2$, are given
by Hermite polynomials type wave functions $\psi_n(x)$, whose
corresponding eigenvalues are given by $E_n = 2 n + 1$, $n= 0, 1, 2, ...$
The Infeld-Hull results are, in this case, $\epsilon = 1$ and $\beta_p(x)
= x$, hence we have $a= d/dx + x$, and equation (\ref{8}) can be rewritten
as
\be
H = -\frac{d^2}{dx^2} + x^2 = a a^{\dagger} -1.
\label{9}
\ee
It must be clear that, in this case, equation (\ref{8}) does not give any
new Hamiltonian. Now, by using (\ref{9}) and the corresponding
factorization (\ref{6}), we get the well known commutation rules
satisfied by the operators $a$, $a^{\dagger}$ and $H$:
\be
[a, a^{\dagger}] = 2, \quad \quad [H, a] = -2 a, \quad \quad
[H, a^{\dagger}] = 2 a^{\dagger}. 
\label{10}
\ee
Furthermore, it is well known that, for the harmonic oscillator problem,
the operators of factorization $a^{\dagger}$ and $a$ are also the
corresponding raising and lowering operators for the eigenfunctions of the
Hamiltonian.

On the other hand, the method by itself gives us the chance of
constructing a first solution to (\ref{3}). In order to make clear this
argument let us consider the equation (\ref{6}), and suppose that there
exists a function $\psi_N(x)$, such that $a\psi_N(x) =0$, hence we will
have $H \psi_N(x) = \epsilon \psi_N(x)$.  The solution to this first order
differential equation, providing the solution to (\ref{7}) as given, is
\be
\psi_N(x) \propto  e^{-\int^x \beta_p(y) dy}.
\label{11}
\ee
If $\psi_N(x)$ satisfies the normalization condition (\ref{2}), then it is
a wave function with eigenvalue $\epsilon$, else it is not a {\it
physical} solution to the Schr\"{o}dinger equation (\ref{3}). For the
system we are dealing with in this section, the function $\psi_N(x)= c_0
\exp(-x^2/2)$ is a physical eigenfunction of $H$ with the eigenvalue
$\epsilon = E_N = 1$, hence $N=0$, and $\psi_N(x)=\psi_0(x)$ is the ground
state wave function of the harmonic oscillator. 

Let us consider now the Hamiltonian (\ref{9}). The solution to the first
order differential equation $a^{\dagger} \psi_M(x) = 0$, given by
$\psi_M(x) \propto \exp(x^2/2)$, is an eigenfunction of $H$ with
eigenvalue $E_M = E_{-1}= -1 \not \in \{ E_n \}$. It is clear that
$\psi_M(x)$ is an {\it unphysical} solution because either it not
satisfies the normalization condition (\ref{2}), or because its eigenvalue
$E_M$ is not allowed in the spectrum of $H$. We want to remark again that
these unphysical solutions are usually discarded by the traditional
methods of solution in Quantum Mechanics. In Section 4 we will have
opportunity to show their relevance in the construction of new exactly
solvable potentials. 

%%%%%%%%%%%%%%%%%%%%%%%%%%%%%%%%%%%%%%%%%%%%%%%%%%%%%
\subsection{The hydrogen-like potential revisited}

The standard procedure to deal with hydrogen-like potentials in QM reduces
to solve the eigenproblem for a particle in a one-dimensional effective
potential $V_l(r)= l(l+1)/r^2 - 2/r$, where $l=0,1,2,...$, is the
azimuthal quantum number and $r$ is a dimensionless radial coordinate. By
simplicity, instead of working with the standard radial wavefunctions
$R_n^l(r)$, we will work with the functions $\psi_{n,l}(r) \equiv r
R_n^l(r)$, with an inner product defined by $\langle \psi_{n,l} ,
\psi_{n,l'} \rangle \equiv 4 \pi \int_{0}^{+\infty} \bar \psi_{n,l}(r)
\psi_{n,l'} (r) dr< \infty$. As it is well known, the eigenvalues of the
radial Hamiltonian
\be
H_l = -\frac{d^2}{dr^2} + \frac{l(l+1)}{r^2} -\frac{2}{r} =
-\frac{d^2}{dr^2} + V_l(r),
\label{13}
\ee
for a fixed $l$, are given by
\be
E_n \equiv E_{l,k} = - \frac{1}{(l+k)^2}; \quad k=1,2,3,...
\label{14}
\ee
where $l+k=n$. Now, for the sake of clarity, let us take the following
notation for the Infeld-Hull factorization operators of the Hamiltonian
(\ref{13}) 
\be
a_l = \frac{d}{dr} + \frac{l}{r} - \frac{1}{l}, \quad \quad a_l^{\dagger}
= -\frac{d}{dr} + \frac{l}{r} - \frac{1}{l},
\label{15}
\ee
hence, the corresponding factorizations (\ref{6}) and (\ref{8}) can be
written as
\be
a_l^{\dagger} a_l - \frac{1}{l^2} = H_l, \quad \quad a_l a_l^{\dagger}-
\frac{1}{l^2} = H_{l-1}. 
\label{16}
\ee
Notice that the right hand side equation (\ref{16}) ({\it e.g.}
(\ref{8})), can be rewritten as
\be
a_{l+1} a_{l+1}^{\dagger}- \frac{1}{(l+1)^2} = H_l.
\label{17}
\ee
In the present case, the operators of factorization $a_l^{\dagger}$ and
$a_l$, map the solutions of the corresponding equation (\ref{3}), with a
given energy, into solutions to (\ref{3}) with the same energy but
changing the value of the azimuthal quantum number $l$. Hence,
$a^{\dagger}_l$ and $a_l$ are respectively the raising and lowering
operators for the quantum number $l$, besides we have $\psi_{n,l+1}(r)
\propto a^{\dagger}_{l+1} \psi_{n,l}(r)$, and $\psi_{n,l-1}(r) \propto
a^{\dagger}_l \psi_{n,l}(r)$.

Let us conclude this section with the observation that the physically
relevant first solution to (\ref{3}), derived from the SDIH method for the
Hamiltonian (\ref{13}), is given by $\psi_M(r) \propto r^{l+1}
\exp(-r/(l+1))$, which is solution of $a_{l+1}^{\dagger} \psi_M(r)=0$. 
Hence, from (\ref{17}) and (\ref{14}), it is apparent that it is an
eigenfunction of $H_l$ with eigenvalue $\epsilon=E_{l,1}= -1/(l+1)^2$. On
the other hand, the {\it unphysical} solution is now given by $\psi_N(r)
\propto r^{-l} \exp(r/(l))$, with the forbidden eigenvalue
$E_{l,0}=-1/l^2$. 

%%%%%%%%%%%%%%%%%%%%%%%%%%%%%%%%%%%%%
\section{Mielnik's factorization}

In this section we shall review the modification of the SDIH method as
introduced by Mielnik. As we have discussed in the introduction, the
Mielnik's method aims to find two operators
\be
A = \frac{d}{dx} + \beta(x), \quad \quad A^{\dagger} = -\frac{d}{dx} +
\beta(x),
\label{18}
\ee
factorizing the Hamiltonian (\ref{5}) in the form
\be
H = A^{\dagger} A + \epsilon,
\label{19}
\ee
where the function $\beta(x)$ is the general solution of the Riccati
equation
\be
-\beta'(x) + \beta^2(x) = V(x) - \epsilon.
\label{20}
\ee
It is well known that, given any particular solution $\beta_p(x)$ to
(\ref{20}), the corresponding general solution is given by two successive
quadratures \cite{I26}, and one can write
\be
\beta(x)  = \beta_p(x)-\frac{d}{dx} \ln \left\{ \lambda- \int e^{ 2 \int^x
\beta_p(y) \, dy } dx \right\},
\label{21}
\ee
where $\lambda$ is an integration constant.

Now, by taking the factorization (\ref{19}) in the opposite order we
get
\be
A A^{\dagger} + \epsilon = H + 2 \beta'_p(x) - 2 \frac{d^2}{dx^2} \ln
\left\{ \lambda- \int e^{ 2 \int^x \beta_p(y) \, dy } dx \right\}.
\label{22}
\ee
From Sections 2.1 and 2.2, it is clear that only the term $ H + 2
\beta'_p(x)$, in the right hand side of (\ref{22}), would correspond to
the initial Hamiltonian $H$. The element of arbitrarness becomes then from
the logarithmic term in such equation. Hence, we can define a new
Hamiltonian $\widetilde H = -(d^2/dx^2) + \widetilde V(x)$, where
\be
\widetilde V(x)  \equiv  V(x) + 2 \beta'(x)
\label{25}
\ee
and such that
\be
\widetilde H = A A^{\dagger} + \epsilon.
\label{23}
\ee
Equations (\ref{25}-\ref{23}) open a new interesting problem: Is the
following eigenvalue equation exactly solvable?
\be
\widetilde H \widetilde \psi(x) = \widetilde E \widetilde \psi(x).
\label{26}
\ee
The answer is affirmative and the corresponding solutions can be constructed
by using the solutions of (\ref{3}), with $H$ given in (\ref{19}), and
the following theorem:

\begin{itemize}
\item[] {\bf Intertwining Theorem}

Let $\psi(x)$ be an eigenfunction of $H$ with eigenvalue $E$, then
$A\psi(x) \neq 0$ is an eigenfunction of $\widetilde H$ with eigenvalue
$E$.  Similarly, if $\widetilde \phi(x)$ is an eigenfunction of
$\widetilde H$ with eigenvalue $\widetilde E$, then $A^{\dagger}
\widetilde \phi(x)  \neq 0$ is an eigenfunction of $H$ with eigenvalue
$\widetilde E$.

{\it Proof.}
\[
\widetilde H [ A \psi(x) ] = A [A^{\dagger} A + \epsilon ] \psi (x) = E [
A \psi(x) ],
\]
\[
H [ A^{\dagger} \widetilde \phi(x) ] = A^{\dagger} [A A^{\dagger} +
\epsilon ] \widetilde \phi(x) = \widetilde E [ A^{\dagger} \widetilde
\phi(x) ]. \quad \quad \mbox{\it Q.E.D.}
\]
\end{itemize}

Then we have $\widetilde \psi(x) \propto A \psi(x)$, and $\psi(x) \propto
A^{\dagger} \widetilde \phi(x)$. Notice that, in the present approach, the
operators of factorization $A^{\dagger}$ and $A$, are not the
corresponding raising and lowering operators for the eigenfunctions of
either $H$ or $\widetilde H$. Now, as we have discussed in Section 2, if
the solutions to (\ref{26}) are wave functions, then they satisfy the
normalization condition (\ref{2}). Hence, we have
\[
\langle \widetilde \psi, \widetilde \psi' \rangle = \langle A \psi, A
\psi' \rangle = \langle A^{\dagger} A \psi, \psi' \rangle = \langle
(H-\epsilon) \psi, \psi' \rangle = (E-\epsilon)  \langle \psi, \psi'
\rangle 
\]
and, if $\psi(x) \in L^2({\bf R})$, we can take $\widetilde \psi(x)=
(E-\epsilon)^{-1/2} A \psi(x)$. It is interesting to notice that, although
we have used all the eigenfunctions of (\ref{19}) in the construction of
the new set $\{ \widetilde \psi(x)= (E-\epsilon)^{-1/2} A \psi(x) \, \vert
\, \psi(x) \in L^2({\bf R}) \}$, it is not, in general, a complete set. 
Suppose there exists a function $\widetilde \psi_{M \epsilon} (x)$ such
that it is orthogonal to all $\widetilde \psi(x)$, then we will have
\[
\langle \widetilde \psi_{M \epsilon}, \widetilde \psi \rangle =
(E-\epsilon)^{-1/2} \langle \widetilde \psi_{M \epsilon}, A \psi \rangle =
(E-\epsilon)^{-1/2} \langle A^{\dagger} \widetilde \psi_{M \epsilon}, \psi
\rangle = 0. 
\]
Therefore, the solution to the first order differential equation
$A^{\dagger} \widetilde \psi_{M \epsilon} (x)= 0$, represents a possible
missing element in the new basis $\{ \widetilde \psi(x) \}$. Such solution
is given by
\be 
\widetilde \psi_{M \epsilon} (x) \propto e^{\int^x \beta(y) dy},
\label{27}
\ee
and by using (\ref{21}) we can rewrite it as
\be
\widetilde \psi_{M \epsilon} (x) \propto \frac{e^{\int^x \beta_p(y) dy}
}{\lambda - \int^x e^{ 2 \int^y \beta_p (z)  dz } dy } = \frac{ \psi_M(x)
}{ \lambda - \int^x \psi_M^2(y)dy },
\label{27a}
\ee
where $\psi_M(x)$ is solution of $a^{\dagger} \psi_M(x) =0$.

Now, if (\ref{27}) satisfies the normalization condition (\ref{2}), then
it is a physical solution to the eigenvalue problem (\ref{26}), with
eigenvalue $\epsilon$ (see equation (\ref{23})), and it has to be added to
the new basis $\{ \widetilde \psi(x) \}$. In general, the eigenfunctions
of $\widetilde H$ are given by $\{ \widetilde \psi_{M \epsilon} (x) \}
\cup \{ \widetilde \psi (x) \}$. Moreover, the Hamiltonian $\widetilde H$
is almost isospectral to $H$, because its spectrum is the same as the
spectrum of $H$ plus a new level at $\epsilon$: $\{ \widetilde E \} = \{
\epsilon, E \}$. In other words, we have solved the eigenvalue equation
(\ref{26}) by using the intertwining theorem and by considering the
missing state (\ref{27}). 

Let us remark that the eigenproblem connected with $H$ can be first solved
by using the SDIH method and after by the Mielnik approach, which gives no
new solutions for $H$, but allows the construction of new exactly solvable
Hamiltonians $\widetilde H$. Therefore, the Mielnik method is a further
step in the systematic construction of exactly solvable problems in QM.

%%%%%%%%%%%%%%%%%%%%%%%%%%%%%%%%%%%%%%%%%%%%%%%%%%%%%%%%%%%%

\subsection{First order intertwining approach}

As we have dicussed in the last section, the Mielnik's approach allows one
to derive new Hamiltonians $\widetilde H$ departing from the factorization
of a given Hamiltonian $H$. It has been also shown that the eigenfunctions
of $H$ are an adequate point of departure in solving the eigenproblem
connected with $\widetilde H$. Now, a natural question arises: Are the
Hamiltonians $H$ and $\widetilde H$ connected by a more direct relation?
The answer is afirmative and it is clearly stated by the algebraic tool
called the {\it first order intertwining} method \cite{C79}. Now, we are
going to show how the basic results of the intertwining method can be
recovered from the Mielnik's approach. With this aim, let us apply the
operator $A$ to the right of equation (\ref{23}), hence, we get
\be
\widetilde H A = [ A A^{\dagger} + \epsilon ] A = A [ A^{\dagger} A +
\epsilon ] = A H.
\label{28}
\ee

Equation (\ref{28}) means that the Hamiltonians $H$ and $\widetilde H$ are
connected by the operator $A$. Notice that the multiplying order of the
operators plays a fundamental role. Suppose now that $H$ is a Hamiltonian
with known solutions to the corresponding equation (\ref{3}), then
(\ref{28}) immediately leads to the first part of the intertwining
theorem. In the intertwining {\it jargon} it is said that $A$ is an
intertwiner operator because it {\it intertwines} (the eigenfunctions of) 
the operators $H$ and $\widetilde H$.  In this case the Hamiltonian $H$ is
called the initial one, while $\widetilde H$ is the intertwined
Hamiltonian, which is to be determined by the method itself.  On the other
hand, the operator $A^\dagger$ intertwines the Hamiltonians $\widetilde H$
and $H$ in the form: 
\be
H A^{\dagger} = [ A^{\dagger} A + \epsilon ] A^{\dagger} = A^{\dagger}[ A
A^{\dagger} + \epsilon ] = A^{\dagger} \widetilde H.
\label{29}
\ee
In this case, considering the multiplying order of the operators, it is
suitable to take $\widetilde H$ as the initial Hamiltonian, therefore,
equation (\ref{29}) allows one to recover the second part of the
intertwining theorem. On the other hand, it is a matter of substitution to
show that equations (\ref{20}) and (\ref{25}) are sufficient conditions
validating the intertwining relationship (\ref{28}). Moreover, all these
equations lead, in a natural form, to the factorization of the initial
Hamiltonian $H$ with the multiplying order of (\ref{19}). In a similar
form, the sufficient conditions to satisfy (\ref{29}) are given by
\be
\beta'(x) + \beta^2(x) = \widetilde V(x) - \epsilon
\label{30}
\ee
and
\be
V(x) = \widetilde V(x)- 2 \beta'(x),
\label{31}
\ee
permitting the factorization of the Hamiltonian $\widetilde H$ as it has
been made in (\ref{23}).

In the last section we have done the assumptions validating the
intertwining relationship (\ref{28}). Now, we shall proceed in the
opposite way, {\it i.e.}, we will suppose that the eigenvalue problem
(\ref{26}) has been solved by the SDIH method and, therefore, we will
focus on solving the corresponding equation for $H$. With this aim, let us
consider now the function $\alpha(x)$ as a particular solution to
(\ref{30}), then the corresponding general solution is given by
\be
\beta(x)  = \alpha(x)+\frac{d}{dx} \ln \left\{ \gamma+ \int e^{- 2 \int^x
\alpha(y) \, dy } dx \right\},
\label{32}
\ee
where $\gamma$ is an integration constant (compare (\ref{32}) with
(\ref{21})). The corresponding eigenfunctions are given by those derived
from the intertwining theorem, plus the missing state
\be
\psi_{N \epsilon} (x) \propto e^{-\int^x \beta(y) dy} = \frac{e^{-\int^x
\alpha(y) dy} }{\gamma + \int^x e^{ -2 \int^y \alpha (z) dz } dy }. 
\label{33}
\ee

Let us remark that, although it is no matter what kind of factorization
({\it e.g.} equations (\ref{19}) or (\ref{23})) we use for the initial
Hamiltonian in the intertwining approach (\ref{28}-\ref{29}), it is
usually important to choose it adequately, in order to get a {\it
physically} permissible intertwined Hamiltonian. Last statements will be
clear in the next subsection, where we are going to apply the Mielnik
method to the harmonic oscillator and to the hydrogen-like potentials by
using the intertwining approach.

%%%%%%%%%%%%%%%%%%%%%%%%%%%%%%%%%%%%%%%%%%%%%%%%%
\subsubsection{Classical isospectral potentials}

\noindent
{\it Harmonic oscillator}

\noindent
We take the initial Hamiltonian $H= -d^2/dx^2 + x^2 + 2$ as it has
been factorized in (\ref{8}-\ref{9}). We want to factorize it in the form
(\ref{23}), hence, by making $ \widetilde V(x) =  x^2 +2 $ in
(\ref{30}), we get the particular solution $\alpha(x) = x$ (see the table
of Section 2). Therefore, the corresponding intertwined potential
(\ref{31}) is given by
\be
V(x)  = x^2 - 2 \frac{d}{dx} \left\{ \frac{e^{-x^2}}{\gamma+ \int_0^x
e^{-y^2} dy} \right\}.
\label{34}
\ee
If $\vert \gamma \vert > \sqrt{\pi}/2$, the above potential has no
singularity and behaves like $x^2$ for $x \rightarrow \pm \infty$; and so,
one obtains here a {\it one}-parameter family of self-adjoint Hamiltonians
$H$ in $L^2(\bf{R})$ \cite{M84}. The first eigenfunctions of $H$ are given
by the intertwining theorem, while the corresponding missing state
(\ref{33}) is given by
\be
\psi_{N \epsilon} (x) \propto  \frac{e^{-x^2/2}}{\gamma+ \int_0^x
e^{-y^2} dy},
\label{35}
\ee
which is a square-integrable function because its behavior for $x
\rightarrow \pm \infty$, provided that $\vert \gamma \vert >
\sqrt{\pi}/2$.\\[2.5ex]
\noindent
{\it Hydrogen-like potential}

\noindent
Let us take $H_l$ as it has been factorized in (\ref{16}). We are going to
factorize it in the form (\ref{19}). By substituting $V_l (r)$ in
(\ref{20}) we get the particular solution $\beta_p(r) = l/r -1/l$, with
fixed $l$ (see the table in Section 2), and the corresponding intertwined
potential $\widetilde V_{l-1}(r)$ (see equation (\ref{25})) is given by
\be
\widetilde V_{l-1}(r) = -\frac{2}{r} + \frac{l(l+1)}{r^2} +2 \frac{d}{dr}
\left\{ \frac{r^{2l} e^{-2r/l}}{\lambda_l - \int_0^r y^{2l} e^{-2 y/l} dy}
\right\}, \quad l \geq 1. 
\label{36}
\ee
If $\lambda_l > (2l)! \, (l/2)^{2l+1}$, or $\lambda_l < 0$, for a fixed
$l$, the third term has no singularities. Furthermore, for $r\rightarrow +
\infty$, we have $\widetilde V_{l-1}(r)= V_{l-1}(r)$, and therefore we
obtain a {\it one}-parameter family of self-adjoint Hamiltonians
$\widetilde H_{l-1}$ in $L^2({\bf R})$ \cite{F84}. The first
eigenfunctions of $\widetilde H_{l-1}$ are given by the intertwining
theorem, while the corresponding missing state (\ref{27a}) is given by
\be
\psi_{M \epsilon} (x) \propto  \frac{r^le^{-r/l}}{\lambda_l - \int_0^r
y^{2l} e^{-2y/l} dy},
\label{37}
\ee
which is a square-integrable function because its behavior at $r=0$, and
in the limit $r \rightarrow +\infty$, provided that $\lambda_l > (2l)! \,
(l/2)^{2l+1}$, or $\lambda_l < 0$.

%%%%%%%%%%%%%%%%%%%%%%%%%%%%%%%%%%%%%%%%%%%%%%
\section{Further factorizations}

In the last sections we have done a survey of the factorization methods
available for the construction of analytical solutions to the
Schr\"{o}dinger equation. Now, we remark that, although the Mielnik method
is a powerful tool in the derivation of new Hamiltonians whose
corresponding eigenproblem is analytically solvable, not all the
particular solutions to the Riccati equation reported in \cite{IH51} have
been taken into account for their generalization through the Mielnik's
approach. On the other hand, it is a matter of fact that the Infeld-Hull
results in the table of Section 2 ({\it e.g.} \cite{IH51}) correspond to
only one factorization energy $\epsilon$ for each potential. Therefore,
there is only one SDIH factorization available for the corresponding
Hamiltonian and we have at hand only one family of isospectral potentials
connected with it.  In the present section we are going to sketch some
steps allowing the selection of new factorization energies. With this aim,
we shall take $H$ in (\ref{19}) as the initial Hamiltonian \cite{comenta}.
The transformation
\be
\beta(x) = -\frac{d}{dx} \ln u(x),
\label{38}
\ee
leads (\ref{20}) to the second order differential equation \cite{R98}
\be
\left[ - \frac{d^2}{dx^2} + V(x) \right] u(x) = \epsilon u(x).
\label{39}
\ee

Hence, the $\beta$-functions are directly connected with the
eigenfunctions of the initial Hamiltonian $H$, corresponding to the
eigenvalue $\epsilon$. Notice that, if the factorization energy $\epsilon$
belongs to the spectrum $\{ E \}$, then the solutions $u(x)$ to (\ref{39})
are the physically relevant functions $\psi(x)$. Now, from Section 3.1.1,
we know that the factorization energy $\epsilon= 1$, for the harmonic
oscillator $H+2$, does not belong to $E_{n+1}$.  On the other hand, for
the hydrogen-like potential we have $\epsilon = -1/l^2$, which is not
allowed in $E_{l,K}$, $K >0$.  Hence, following \cite{M84} and \cite{F84},
we shall consider just the cases where the factorization energy $\epsilon$
does not belong to the spectrum of $H$. Therefore, the solutions to
(\ref{39}) do not have direct physical meaning, but we have just seen that
they naturally lead to the factorization of the Hamiltonians $H$ and
$\widetilde H$ in the spirit of Mielnik's approach, providing as well the
explicit form for the new potentials $\widetilde V(x)$.

In order to solve (\ref{39}) we take the factorization of $H$ as given in
(\ref{6}), hence we have two options:
\begin{itemize}
\item[]
({\it i\/}) $a\, u(x) =0$,
\item[]
({\it ii\/}) $a\, u(x) = w(x) \neq 0$; \quad $a^{\dagger} a\, u(x) = 0 \,
\Rightarrow \, a^{\dagger}\, w(x) = 0$. 
\end{itemize}
Notice that $u(x)= \psi_N(x)$, with $\psi_N(x)$ given in (\ref{11}), is
the solution of ({\it i\/}), whereas $w(x) = \psi_M(x)= c_1/\psi_N(x)$,
with $c_1$ a constant, is the solution to the right hand side equation
({\it ii\/}). Hence, the general solution to the left hand side equation
({\it ii\/})  is given by
\be
u(x) = \psi_N(x) \left[ c_0 + c_1 \int^x \psi_N^{-2} (y) \, dy
\right],
\label{40}
\ee
where $c_0$ is an integration constant. Now, by using (\ref{38}) and
(\ref{11}), with $c_1 = -1$ and $c_0 = \lambda$, we get (\ref{21}). The
general solution of (\ref{20}) is then constructed by the general solution
of (\ref{39}), with the {\it unphysical} eigenvalue $\epsilon$. 

%%%%%%%%%%%%%%%%%%%%%%%%%%%%%%%%%%%%%%%%%%%%%%%%%%
\subsection{New isospectral potentials}

In this section we are going to solve, explicitly, the eigenvalue problem
(\ref{39}) for the harmonic oscillator and the hydrogen-like potentials.
Hence, our results will bring out new factorization energies permitting
the construction of new one-parameter families of potentials isospectral
to each one of the potentials mentioned before.

%%%%%%%%%%%%%%%%%%%%%%%%%%%%%%%%%%%%%%%%%%%%%%%%%%%
\subsubsection{The hydrogen-like potential}

In order to solve (\ref{39}), with $V(x) = V_l(r)$, let us write the
factorization energy as $\epsilon^{(k)}_l \equiv -1/(l+ k)^2$, $k \neq
K$, $l>0$. It is clear that, for a fixed value of $l$, we have
$\epsilon^{(k)}_l \neq E_{l,K}$, $\forall \, k \neq K$. Now we make the
transformation \cite{R98}
\be
u^{(k)}_l(r) = r^{-l} e^{r/(l+k)} \Phi^{(k)}_l(r),
\label{41}
\ee 
leading to a confluent hypergeometric equation for $\Phi^{(k)}_l(r)$,
whose general solution for the discrete values $k=0,-1,-2,...,-(l-1)$ is
given by the linear combination of confluent hypergeometric functions
\cite{WG89}
\bea
\nonumber
\Phi^{(k)}_l(r) &=& {}_1F_1[k,-2l,-2r/(l+k)]\\
\nonumber
&-& \nu_{lk}
[2r/(l+k)]^{1+2l}
{}_1F_1 [1+k+2l,2+2l,-2r/(l+k)],
\eea
with
\[
\nu_{lk} \equiv \frac{\Gamma (1 + \vert k \vert )}{\Gamma (2 l + 2)}
\frac{ \Gamma (-2l)  } {\Gamma (-2l + \vert k\vert )} \, \lambda^{(k)}_l ,
\quad \quad l>0.
\]
The general solution to the corresponding Riccati equation (\ref{20})
arises after introducing (\ref{41}) in (\ref{38}), which gives:
\[
\beta^{(k)}_l(r) = \frac{l}{r} - \frac{1}{(l+k)} - \frac{d}{dr} \ln
\Phi^{(k)}_l(r), \quad \quad l>0.
\]
The intertwined potentials $\widetilde V^{(k)}_l(r)$ have the same
singularity at $r=0$ as $V_{l-1}(r)$, provided that $\lambda^{(k)}_l$
takes only values in the domain \cite{R98}:
\[
\lambda^{(k)}_l \in \cases{ (-\infty, 1), & for $\vert k \vert$ even;\cr
\cr
(1, \infty), & for $\vert k \vert$ odd. \cr}
\]
By taking $k=0$ and $\lambda_l^{(0)} = 0$, the last results reduce to the
corresponding SDIH results \cite{IH51}, for the same value of $k$ and
$\lambda^{(0)}_l= (2l)! ( l/2)^{2l+1} \gamma^{-1}_l$, we get the
Fern\'andez results \cite{F84} and, finally, the corresponding Abraham and
Moses results \cite{A80} arise for $k=0$ and $\lambda^{(0)}_1 \rightarrow
1$, with $l=1$.

%%%%%%%%%%%%%%%%%%%%%%%%%%%%%%%%%%%%%%%%%%%%%%%%%%%%
\subsubsection{The harmonic oscillator}

For the harmonic oscillator potential we take $\widetilde V(x) = x^2$ as
the initial potential, hence, in order to solve the corresponding equation
(\ref{39}), let us make the transformation
\be
\widetilde u(x) = \Phi(x) \exp (-x^2/2),
\label{42}
\ee
and by introducing instead of $x$ the variable $y=x^2$ we then find
$\Phi(y)$ to satisfy a confluent hypergeometric type differential
equation, whose general solution is given by \cite{WG89}
\[
\Phi(y) = {}_1 F_1 ( \mbox{ $\frac{1-\epsilon}{4}, \frac{1}{2}; 
y$} ) + \nu \mbox{ $\frac{\Gamma( \mbox{ $\frac{3-\epsilon}{4}$} )} {
\Gamma( \mbox{ $\frac{1-\epsilon}{4}$} )}$ } y^{1/2} {}_1 F_1 (\mbox{
$\frac{3-\epsilon}{4}, \frac{3}{2}, y$}). 
\]
The general solution to the corresponding Riccati equation (\ref{30})
arises after the transformation $\beta(x) = (d/dx) \ln \widetilde u(x)$,
which gives
\[
\beta(x)= -x - \frac{d}{dx} \ln \Phi(x^2).
\]
The intertwined potential becomes free of singularities if $\epsilon < 1$
and $\vert \nu \vert <1$ \cite{F97,J97}. In particular, we can take
$\epsilon^{(k)} = -2k-1$, with $k=0,1,2,..$.\cite{F97}.

In this case, the SDIH results arise by taking $k=0$ and $\nu = 0$, the
Mielnik results \cite{M84} are recovered with $k=0$ and $\nu^{(0)} =
(\sqrt{\pi}/2)  \gamma^{-1}$ and, by taking the same value of $k$, with
$\nu \rightarrow 1$, the Abraham and Moses family is recovered. We
conclude this Section remarking that the factorization energies
$\epsilon^{(k)}_l$, and $\epsilon^{(k)}$, generalize the choice made for
the SDIH and Mielnik factorizations of the hydrogen-like and harmonic
oscillators potentials respectively.

%%%%%%%%%%%%%%%%%%%%%%%%%%%%%%%%%%%%%%%%%%%%%%%%%%%%%%%
\subsection{Higher order factorizations}

The Mielnik factorization allows us to construct further new families of
exactly solvable potentials departing from any exactly solvable problem
and iterating after the procedure. The new families will inherit the
number of parameters from the old ones. The keystone of that construction
is given by \cite{F97,R98,M98}: 
\[
\beta_{n+1}(x) = -\beta_n(x) - \left( \frac{\epsilon_{n+1}
-\epsilon_n}{\beta_{n+1} -\beta_n} \right), \quad n=1,2,...
\]
where $n$ denotes the number of iterations.

Notice that the last equation works immediately for $\epsilon_{n+1} \neq
\epsilon_n$. Hence, by giving $n$ different factorization energies one can
iterate $n$ times the Mielnik's approach constructing $n$th-parametric
families of analytically solvable potentials. The results presented in
this paper reproduce those recently derived in \cite{F97,R98}. Although we
have shown how this method works by solving the discrete problems
connected whith two interesting physical systems, its applications include
the study of systems with continuous spectrum ({\it e.g.} $\epsilon_{n+1}
\rightarrow \epsilon_n$) \cite{M98}.

%%%%%%%%%%%%%%%%% ACKNOWLEDGMENTS %%%%%%%%%%%
{\section*{Acknowledgments}}

\noindent 
I want to take full advantage of this opportunity to thank Prof. 
D.~J.~Fern\'andez, who drew my attention to this very interesting field. I
am also grateful to L.~M. Nieto and J.~Negro for helpful discussions and
to the Organizing Committee of the First International Workshop on
``SYMMETRIES IN QUANTUM MECHANICS AND QUANTUM OPTICS'' for financial
support. This work has been supported by CONACyT, M\'exico (91799, Ref
973035), and partially supported by Junta de Castilla y Le\'on, Spain
(project C02/97). The kind hospitality at Departamento de F\'{\i}sica
Te\'orica is also aknowledged.

\bigskip

%%%%%%%%%%%%%%%%%%%%%%%%%%%%%%%%%%%%%%%%%%%%%%%%%%%%%%

%\section*{References}


\begin{thebibliography}{99}
\bibitem{comenta}
It is clear that the results and discussions presented in this section for
$H$ will be valid for $\widetilde H$ by taking the corresponding changes.
\end{thebibliography}
\end{document}